\documentclass[a4paper,twocolumn,11pt]{quantumarticle}
\pdfoutput=1
\usepackage[utf8]{inputenc}
\usepackage[english]{babel}
\usepackage[T1]{fontenc}
\usepackage{amsmath}
\usepackage{amssymb}
\usepackage{hyperref}

\usepackage{tikz}
\usepackage{lipsum}

\usepackage{graphicx}
\usepackage{xcolor}
\usepackage{dcolumn}
\usepackage{bm}
\usepackage{amsthm}
\usepackage{algorithm,algcompatible}

\usepackage{caption}
\DeclareCaptionFormat{myformat}{#3}
\usepackage[noend]{algpseudocode}
\usepackage{algpseudocode}

\usepackage{braket}
\usepackage{ctable}
\usepackage{tabularx}
\usepackage{array}
\usepackage{float}
\usepackage{hyperref}

\usepackage[numbers]{natbib}
\begin{document}

\title{A hybrid quantum-classical approach to warm-starting optimization}

\author{Vanessa Dehn}
\email{vanessa.dehn@iaf.fraunhofer.de}
\author{Thomas Wellens}
\email{thomas.wellens@iaf.fraunhofer.de}
\affiliation{Fraunhofer Institute for Applied Solid State Physics IAF, Tullastr. 72, 79108 Freiburg, Germany}

\maketitle

\begin{abstract}
The Quantum Approximate Optimization Algorithm (QAOA) is a promising candidate for solving combinatorial optimization problems more efficiently than classical computers. Recent studies have shown that warm-starting the standard algorithm improves the performance. In this paper we compare the performance of standard QAOA with that of warm-start QAOA in the context of portfolio optimization and investigate the warm-start approach for different problem instances. 
In particular, we analyze the extent to which the improved performance of warm-start QAOA is due to quantum effects, and show that the results can be reproduced or even surpassed by a purely classical preprocessing of the original problem followed by standard QAOA.
\end{abstract}

\section{\label{sec:Intro}Introduction}
The quantum approximate optimization algorithm (QAOA) \cite{farhi2014quantum,Farhi2022quantumapproximate}
is often presented as a candidate for the efficient solution of combinatorial optimization problems in the current noisy intermediate-scale quantum (NISQ) era \cite{grange2022introduction}, i.e. the era of working on quantum hardware with error rates and limitations in size \cite{Lau2022}. However, an advantage over classical algorithms has not yet been proven, so various modifications of QAOA such as warm-start QAOA, recursive QAOA, spanning tree QAOA \cite{Egger2021warmstartingquantum, bae2023recursive, Bravyi_2020, Wurtz2021} or using an alternative objective function \cite{PhysRevResearch.4.023225} are proposed to further improve its performance.

The warm-start approach is a well-known recipe for reducing the time needed to solve an optimization problem by starting the optimization with an efficiently computable approximate solution \cite{poloczek2016warm}. Applying this concept to quantum optimization, it has been shown that warm-starting the QAOA based on the classically obtained solution of a relaxed continuous optimization problem shows an improvement at a lower depth \cite{Egger2021warmstartingquantum, tate2022bridging}. The warm-start QAOA has been applied and discussed in the context of the MaxCut problem \cite{ 10.1145/227683.227684,tate2022warmstarted}, where the key idea is to find the maximum cut of a graph. Various problems, such as unstructured clustering problems, can be mapped to such a graph optimization problem and are shown to be solved using the warm-start QAOA \cite{beaulieu2021maxcut}. An  extensive study of warm-start including the selection and optimization of hyperparameters (also applied to the MaxCut problem) is reported in \cite{electronics11071033}. In addition to the above mentioned versions of warm-start, referred to as the continuous warm-start
procedure, Egger et al. \cite{Egger2021warmstartingquantum} have proposed modifications such as the rounded warm-start QAOA, where the initial state is generated by randomly rounding the SDP (semidefinite programming) relaxation of the QUBO problem. Another approach, called classically-boosted quantum optimization algorithm (CBQOA) \cite{wang2022classicallyboosted}, also uses a rounded solution of the SDP relaxation as initial state, followed by an efficiently-implementable continuous-time quantum walk. 

From a general perspective, warm-starting quantum optimization is an example of a hybrid approach, where classical tools for solving a combinatorial optimization problem are combined with quantum methods. The idea is to analyze the processing steps throughout an optimization algorithm and to evaluate which steps are more efficient with classical methods and in which steps quantum methods are more suitable \cite{Weigold2021_HybridPatterns,mcclean21}.
In this paper, we focus on the problem of portfolio optimization and investigate the original warm-start QAOA proposed in \cite{Egger2021warmstartingquantum} to analyze what factors are responsible for its improved performance compared to its standard version. In particular, we compare the performance of warm-start QAOA and standard QAOA for different problem instances, depending on how well the classically obtained solution of the relaxed problem (serving as the starting point of warm-start QAOA) agrees with the desired solution of the combinatorial optimization problem. Moreover, we
propose a classical preprocessing scheme for standard QAOA, thereby formulating a hybrid quantum-classical approach to warm-start optimization that reproduces or even outperforms the performance of the actual warm-start QAOA. 

This paper is organized as follows: in Sec. \ref{sec:Background}, we give a short introduction to the portfolio optimization problem and a brief presentation of solving the problem for random problem instances by using the warm-start QAOA in comparison to the standard QAOA. We then subdivide our problem instances into "hot" and "cold" optimizable instances, i.e. instances with a relaxed solution that is either closer to or further from the optimal solution, and discuss the performance of warm-start vs. standard QAOA for different instances in Sec. \ref{sec:warm_start_instances}. In Sec. \ref{sec:Classicalpreprocessing}, we present a classical preprocessing method for the standard QAOA routine and analyse its performance in comparison with standard and warm-start QAOA for random, hot and cold problem instances. Finally, in Sec.\ref{sec:Conclusion} we conclude and give a brief perspective for further investigations.

\section{\label{sec:Background}Background}
\subsection{\label{sec:Portfolioproblem}Formulation of the Portfolio Problem}
In the general quadratic unconstrained binary optimization (QUBO) problem, a cost function is defined on $N$ 
binary 
variables
$F:\mathbb{B}^N \rightarrow \mathbb{R}$:
\begin{equation}
    F({\bf x}) = \sum_{i,j=1}^{N}F_{ij}x_i x_j + \sum_{i=1}^{N} f_i x_i~,
    \label{eq:qubo}
\end{equation}
with symmetric matrix $F_{ij} \in \mathbb{R}^{N\times N}$, vector $f_i \in \mathbb{R}^N$ and $N$ binary variables ${\bf x}=(x_1,x_2,\dots,x_N) \in \{0,1\}^N$. 
Note that, since $x_i = x_i^2$ for binary variables, the vector $f_i$ can also be added to the diagonal of the matrix $F_{ij}$. However, we will keep the linear term in the following, since it will be relevant for the continuous relaxation (see below).
The solution of the problem is given by $\bf x^{\rm opt}$ that minimizes the above cost function.

Considering the portfolio problem as our problem model, we can directly identify the matrix $F_{ij}$ with the covariance matrix of the stock returns $\sigma_{ij}$ and the vector $f_i$ with the expected return $\mu_i$, respectively. The binary variables $x_i$ represent the portfolio weights, which are 1 or 0, depending on the stock being chosen for the portfolio or not. Additionally, with the parameter $q \in [0,1]$ the risk-preference, depending on whether the risk or the return is to be taken into account with larger weight, can be set. The function that has to be minimized thus reads:\\
\begin{equation}
    F_C({\bf x})=q\sum_{i,j=1}^{N}x_i x_j \sigma_{ij}-(1-q)\sum_{i=1}^{N} x_i \mu_i~.
    \label{eq:costfunction}
\end{equation}
The investment of a fixed amount of money can be addressed by introducing a budget constraint $B=\sum_{i}x_i$, where $B$ is the number of the assets selected from the $N$ available assets for the portfolio.
In order to deal with this constraint, a penalty term, denoted with $A$, is added to the cost function. Finally, the cost function of our QUBO problem is obtained as:
\begin{equation}
   F({\bf x}) = F_C({\bf x}) + A\left(B-\sum_{i=1}^N x_i\right)^2~.
   \label{eq:costfunction_portfolio}
\end{equation}
The routine that is used to determine a suitable factor for $A$ is described in \cite{Brandhofer2022}.
In the following, we refer to a created portfolio as "feasible" only if the budget constraint is met.
\subsection{\label{sec:standard and warm-start}Standard and warm-start QAOA}

\begin{figure*}[ht!]
\centering
\includegraphics[width=0.95\textwidth]{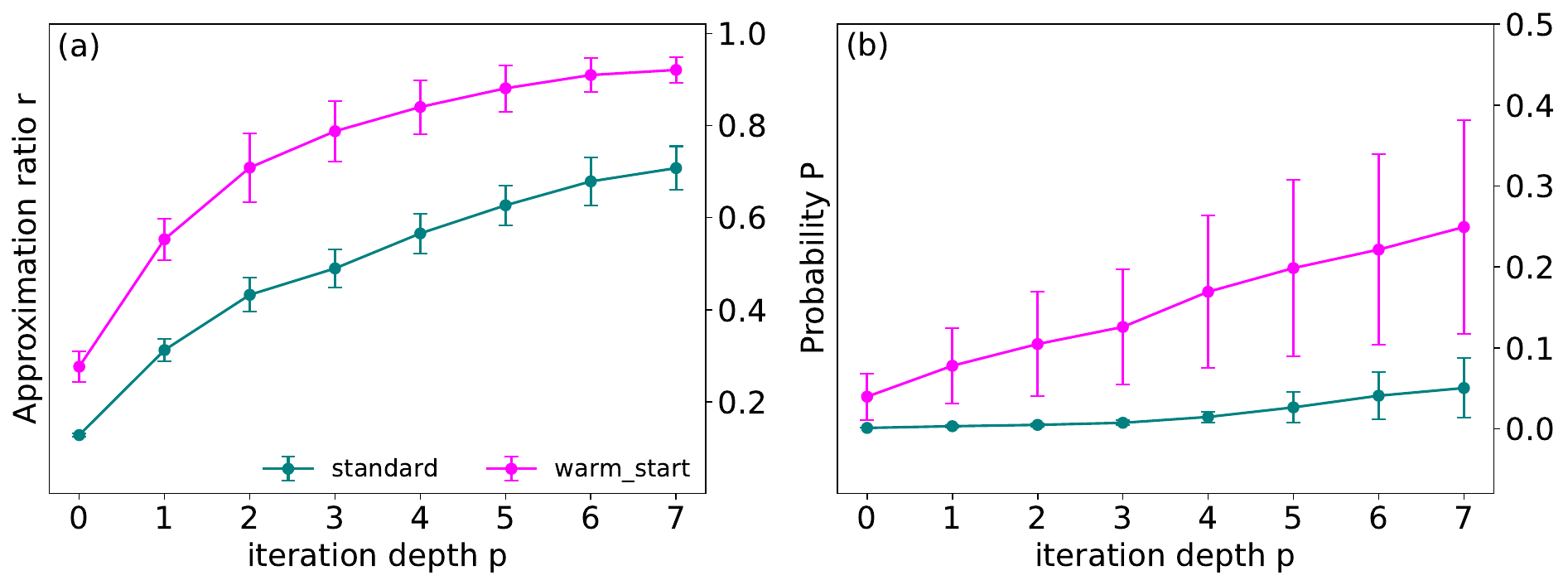}
\captionsetup{justification=raggedright}
\caption{Mean approximation ratio $r$ (a) and ground state probability $P$ (b), both as a function of the number $p$ of QAOA iterations for standard (teal line) and warm-start (pink line) QAOA for an ensemble of 20 random portfolio instances consisting of $N=10$ assets with budget constraint of $B=5$. Compared to the standard mixer, the warm-start mixer yields a better performance (i.e. approximation ratio tends towards $r=1$ and probabilities are higher) since the initial state ($p=0$) for the warm-start version, prepared with the relaxed solution, is already closer to the optimal solution than for the standard mixer.}
\label{fig:ws_std_approx}
\end{figure*}

In order to address the optimization problem with the quantum computer, the cost function has to be mapped onto a cost Hamiltonian $\hat{F}$. This can be done by converting the binary variables to operators with $x_i = (\hat{I}_i + \hat{Z}_i/2)$, with $\hat{I}_i~\text{and}~\hat{Z}_i$ being the identity and the Pauli-$\hat{Z}$ operator acting on qubit $i$ ($i=1,2,\dots,N$).
The QAOA circuit then generates the parametrized variational quantum state
\begin{eqnarray}
\label{eq:quantum state}
    \ket{\psi_{\gamma,\beta}}_{\text{std}} = \hat{U}_{\text{std}}(\beta_p)e^{-i\gamma_p \hat{F}} ...~\hat{U}_{\text{std}}(\beta_1)e^{-i\gamma_1 \hat{F}}\ket{\psi_0}_{\text{std}}
\end{eqnarray}
with parameters $\vec{\gamma}=(\gamma_1,...,\gamma_p)$ and $\vec{\beta}=(\beta_1,...,\beta_p)$ and number of iterations $p$, and standard mixer
$\hat{U}_{\text{std}}$:
\begin{equation}
    \hat{U}_{\text{std}}(\beta) = e^{i\beta\sum_{i=1}^{N}\hat{X}_i} .
\end{equation}
The initial state is chosen to be $\ket{\psi_0}=\ket{+}^{\otimes N}$, which is the minimum energy eigenstate of the mixing operator $-\sum_i \hat{X}_i$. Further, all qubits are measured in the computational basis to determine the expectation value $\langle \hat{F} \rangle$. This intermediate result is then passed on to a classical optimizer, which updates the parameters to minimize the expectation value. \\
The form of the generated quantum state of QAOA, see Eq.~\eqref{eq:quantum state}, is inspired by adiabatic quantum computing (AQC) in terms of starting in the ground state of the mixing Hamiltonian, which is then gradually transferred to the ground state of the cost Hamiltonian by approximating the adiabatic annealing path via Trotterization \cite{Sack2021quantumannealing,fratus2022describing} for iteration depth $p \rightarrow \infty$. Therefore, the performance of QAOA improves with increasing $p$ \cite{zhou2020}.

In contrast to the standard variant of QAOA, which starts from a uniform superposition $\ket{\psi_0}=\ket{+}^{\otimes N}$ of all portfolios,
the warm-start variant, introduced by \cite{Egger2021warmstartingquantum, tate2022warmstarted}, starts with solving the continuous relaxation of the QUBO problem
\begin{equation}
    {\bf x^*}=\underset{{\bf \tilde{x}}\in [0,1]^N}{\text{arg min}}~F({\bf \tilde{x}})
    \label{eq:relax}
\end{equation}
where the variables ${\bf \tilde{x}}$ are not binary, but real numbers $\in [0,1]$. The continuous optimization problem \eqref{eq:relax} can be easily solved by classical optimization if the matrix $F_{ij}$, see Eq.~\eqref{eq:qubo}, is positive-semidefinite (leading to a convex quadratic problem).
For our problem, this is the case, since
$F_{ij}=q\sigma_{ij}+A\delta_{ij}$, see Eqs.~\eqref{eq:qubo}-\eqref{eq:costfunction_portfolio},
with positive semidefinite $\sigma_{ij}$ (as a covariance matrix) and $q,A\geq 0$. In the general case, if $F_{ij}$ is not positive semidefinite, the problem can be "convexified" either by changing the diagonal of $F$ together with the linear term $f_i$ such that the binary problem remains invariant (i.e. $F_{ii}\to F_{ii}+c$, $f_i\to f_i-c$) \cite{Billionnet2007} or by a semidefinite programming (SDP) procedure, see \cite{Egger2021warmstartingquantum}.
 
The optimal solution ${\bf x^*}=(x_1^*,x_2^*,\dots,x_N^*)$ of Eq.~\eqref{eq:relax} can now be used to initialize the QAOA, thus warm-starting it.
For the warm-start algorithm as introduced in~\cite{Egger2021warmstartingquantum}, the initial state of the standard algorithm $\ket{\psi_0}_{\text{std}}=\ket{+}^{\otimes N}$ is replaced by the state 
\begin{eqnarray}
  \ket{\psi_0}_{\text{WS}} = \overset{N}{\underset{i=1}{\otimes}}\hat{R}_Y(\theta_i)\ket{0}_N  
\end{eqnarray}
with angle $\theta_i = 2\,\text{arcsin}(\sqrt{x_i^{*}})$. Further, the mixer $\hat{U}_{\text{std}}(\beta)$ is replaced by $\hat{U}_{\text{ws}}(\beta)=e^{-i\beta \hat{M}_\text{ws}}$ with mixing operator
\begin{eqnarray}
 \hat{M}_{\text{ws}}= -\sum_{i=1}^N \left[\text{sin}(\theta_i)\hat{X}_i+\text{cos}(\theta_i)\hat{Z}_i\right]
\end{eqnarray}
whose ground state is the initial state $\ket{{\psi_0}}_{\text{WS}}$. When a bit in the relaxed solution is exactly set to 0 or 1 (i.e. $x_i^*=0$ or $x_i^*=1$), the respective qubit is then initialized either in the $\ket{0}$ or $\ket{1}$ state. Since the cost Hamiltonian only applies $Z$-gates, the qubits in the above mentioned states will remain in these states during the optimization. To avoid this, a regularization parameter can be introduced \cite{Egger2021warmstartingquantum} which, however, we will not consider in the present paper.

To evaluate the performance of the optimization procedure, i.e. the quality of the solution, we use as a measure for both variants (standard and warm-start) the approximation ratio 
\begin{equation}
    r(x_1,...,x_N) = \begin{cases}
    \frac{F_C(x_1,...,x_N)-F_C^{\text{max}}}{F_C^{\text{min}}-F_C^{\text{max}}} &\text{if $\sum_i x_i = B$}\\
    0 &\text{if $\sum_i x_i \neq B$} \\
    \end{cases}
\end{equation}
averaged over the different measurements results $x_1,...,x_N$ performed on the final state of the optimized QAOA circuit,
where $F_C^{\text{max}}$ and $F_C^{\text{min}}$ indicating the worst and the best solution among all feasible solutions. Measuring the optimal solution yields an approximation ratio of 1, whereas measuring the worst feasible or unfeasible solution yields an approximation ratio of 0. Note, that the approximation ratio is formulated with the cost function $F_C$, see Eq.~\eqref{eq:costfunction}, to have a measure that is independent of the choice of the penalty term $A$. As alternative measure, we also consider the probability $P$ of measuring the optimal solution (also called "ground state probability" in the following).

Concerning our portfolio model, we use an ensemble of 20 randomly generated portfolio instances, each consisting of $N=10$ assets (and budget constraint $B=5$), corresponding to $N=10$ qubits in the quantum circuit. The solution of the relaxed problem, which is used to initialize the warm-start QAOA, is obtained using a gradient-based optimizer, whereas a gradient-free optimizer was chosen for the classical subroutine of QAOA (see \cite{Brandhofer2022} for details). The quantum circuits are executed 
using the qasm simulator with a fixed number of shots (here: 1000). 

For both variants (standard and warm-start), we compare the mean approximation ratio $r$ and the ground state probability $P$ for increasing iteration depth $p$ (up to $p_{\text{max}}=7$), see Fig.~\ref{fig:ws_std_approx}. The standard deviation obtained from the 20 random instances is represented by the error bars. 
As expected, we find that, both, the approximation ratio $r$ and the ground state probability $P$ for the warm-start QAOA is significantly higher than for the standard QAOA, since the initial state (at $p=0$) for warm-start QAOA is already closer to the desired ground state energy of the problem Hamiltonian, i.e. the optimal solution. From the varying errorbars, we also observe that the fluctuations between the 20 random instances are larger for warm-start QAOA (especially in case of the ground state probability $P$). Note that, in contrast to standard QAOA, warm-start QAOA exhibits some fluctuations already for the initial state ($p=0$), depending on how well the solutions of the relaxed and the binary problem agree with each other.

\section{\label{sec:warm_start_instances}Warm-start QAOA for different Problem Instances}

\begin{figure}
\centering
\includegraphics[width=0.45\textwidth]
{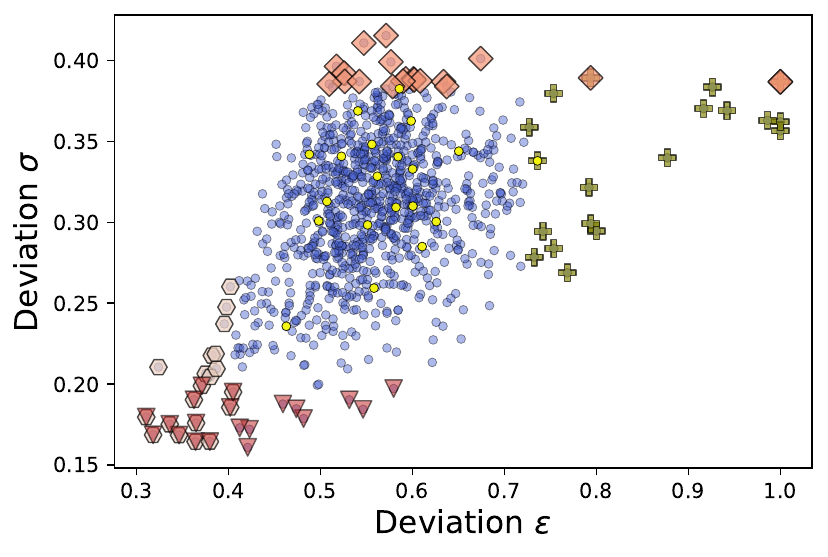}
\captionsetup{justification=raggedright}
\caption{Deviations $\varepsilon$ and $\sigma$ between the solutions of the relaxed and binary problem for 1000 random instances (blue circles). For both measures ($\varepsilon$ and $\sigma$), we identify "hot" and "cold" subsets of 20 instances exhibiting the smallest or largest deviation, respectively ("$\sigma$-hot": red down facing triangles, "$\varepsilon$-hot": tan hexagons, "$\sigma$-cold": orange diamonds, and "$\varepsilon$-cold": olive plus symbols). The yellow circles represent the 20 randomly chosen instances considered in Fig.~\ref{fig:ws_std_approx}.}
\label{fig:scatter}
\end{figure}

To clarify the behaviour for different problem instances,
we generate a larger set consisting of 1000 random portfolio instances, each consisting of $N=10$ assets.
Then, we analyze the distance between the solutions $\bf x^*$ and $\bf x^{\rm opt}$ of the relaxed and the binary problem in order to identify samples of "hot" and "cold" instances where, respectively, the starting point of warm-start QAOA is close (or not close) to the optimal solution of the binary problem.

To quantify the distance, we compare the two vectors $\bf x^*$ and $\bf x^{\rm opt}$ and determine the maximum deviation per instance, denoted with $\varepsilon$:
\begin{equation}
    \varepsilon = \underset{i}{\text{max}}|x_i^{*}-x_i^{\text{opt}}| .
\end{equation}
A large deviation refers to what we call a "cold" instance, whereas for a small deviation the instances are called "hot". 

\begin{figure*}[ht!]
	\centering
        \includegraphics[width=0.95\textwidth]{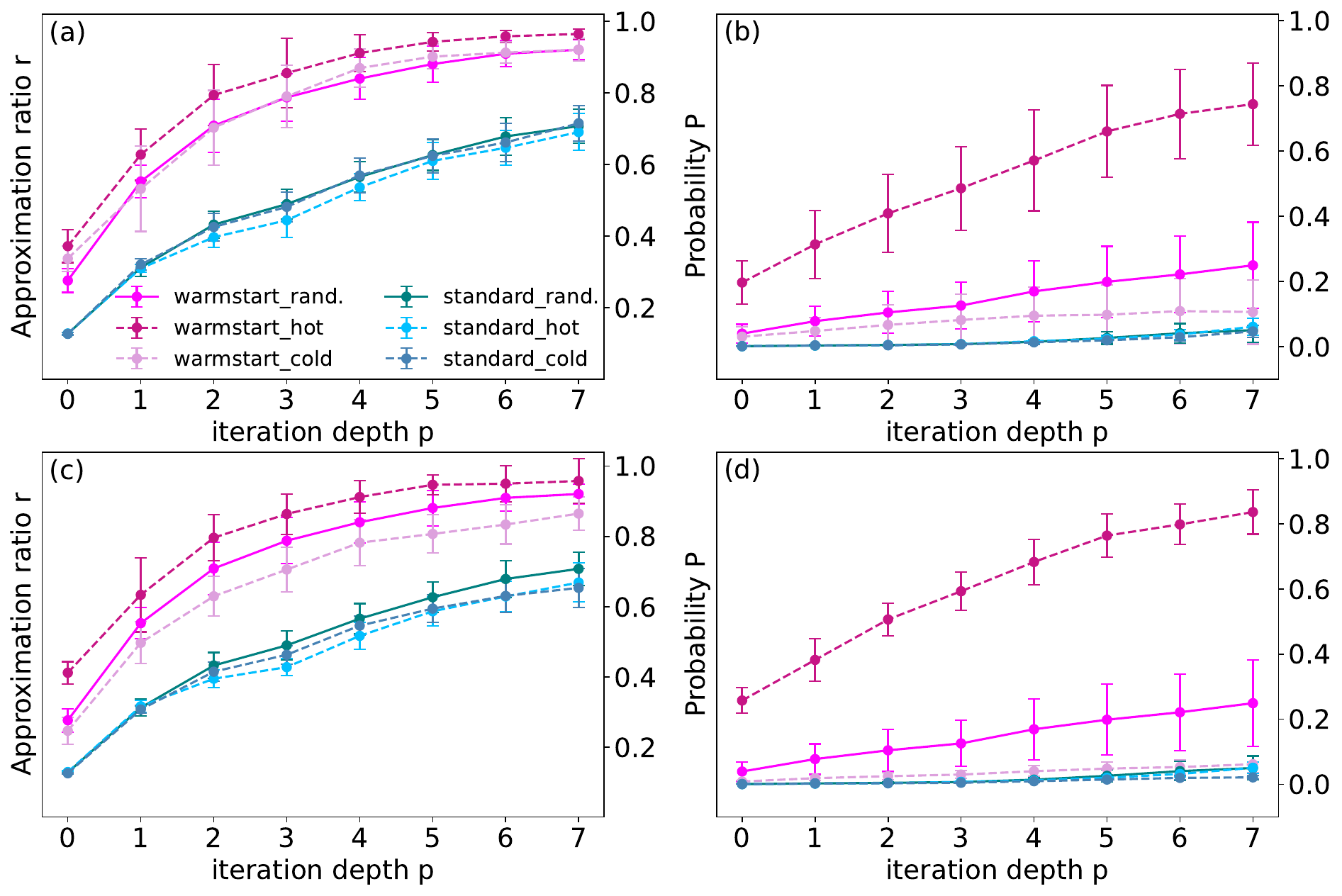}
    \captionsetup{justification=raggedright}
	\caption{(a,b) Same as Fig.~\ref{fig:ws_std_approx}, but additionally for $\sigma$-hot and $\sigma$-cold problem instances. (c,d) Same as Fig.~\ref{fig:ws_std_approx}, but additionally for $\varepsilon$-hot and $\varepsilon$-cold problem instances. For both measures ($\sigma$ and $\varepsilon$), the warm-start performance is either reduced or improved by choosing cold or hot instances. In particular, hot instances exhibit significantly higher values of the ground state probability. The advantage of warm-start QAOA compared to standard QAOA, however, still holds for cold instances. In contrast, the performance of standard QAOA is not sensitive to differentiating between cold or hot instances.}
 	\label{fig:good_bad_eps2}
\end{figure*}

In addition to the maximum deviation $\varepsilon$, we also consider the root  mean square error (RMSE) 
\begin{equation}
    \sigma = \sqrt{\sum_i 
    \left(x_i^*-x_i^{\rm opt}\right)^2/N}
\end{equation}
as a second measure, which takes into account the differences between all pairs of values (not only the maximum).
For both measures, we identify, from our ensemble of 1000 random instances, subsets of 20 instances exhibiting the smallest ("hot") or largest ("cold") deviation, respectively, see Fig.~\ref{fig:scatter}.  

In Fig. \ref{fig:good_bad_eps2} the approximation ratios as well as the ground state probabilities of the "hot" and "cold" instances for both QAOA variants are shown as a function of the iteration depth $p$. For comparison, the values of the randomly chosen instances, which are already displayed in Fig.~\ref{fig:ws_std_approx}(a), are also shown. It is noticeable that, for the standard version, differentiating between the two subsets does not make a difference to the result, neither for the approximation ratio nor for the probability, i.e. all deviations are within the error bars, see Fig.~\ref{fig:good_bad_eps2}(a)-(d). With respect to the warm-start, we see a clear improvement in performance in terms of the approximation ratio when using the "hot" instances (Fig.~\ref{fig:good_bad_eps2}(a,c)). For the ground state probabilities, the increase is even significantly higher (Fig.~\ref{fig:good_bad_eps2}(b,d)). However, the improved performance of warm-start QAOA as compared to standard QAOA is not restricted to hot instances, but still holds for clod instances.

\section{\label{sec:Classicalpreprocessing}Standard QAOA with Classical Preprocessing}

Above, we have seen that warm-start QAOA displays a better performance (both, in terms of approximation ratio and ground state probability) than standard QAOA - especially (but not only) for those problem instances where the classically obtained relaxed solution is close to the optimal solution.
In this chapter, we will show that a similar increase of performance can be obtained by a purely classical preprocessing of the original optimization problem, which can then be solved with standard QAOA.  

\subsubsection{Elimination of Variables}

The solution of the relaxed problem frequently exhibits some bits, which are already exactly set to 0 and 1 (i.e., $x_i^*=0$ or $x_i^*=1$). Since, as already mentioned above, their value is not changed by the warm-start QAOA algorithm, one can as well eliminate those bits from the original problem and run warm-start QAOA on a reduced set of qubits without changing its performance. 

We will follow this idea and extend this procedure also to those bits where the relaxed solution $x_i^*$ is not necessarily exactly 0 or 1, but close to it. For this purpose, we introduce two bounds $\delta_0$ and $\delta_1$ and round the relaxed solution as follows:

\begin{algorithm}[H]
\caption{Rounding Scheme}\label{tab:rounding scheme}
\begin{algorithmic}
\centering
\If{$x_i^* \leq \delta_0$} 
\State $x_i = 0$
\ElsIf{$x_i^* \geq 1-\delta_1$} 
\State $x_i = 1$ 
\EndIf
\end{algorithmic}
\end{algorithm}

Depending on the choice of $\delta_0$ and $\delta_1$, a certain subset of variables $x_i$ is thereby fixed to either 0 or 1. We then look at the reduced problem depending only on those variables that have not been rounded, which we finally solve using standard QAOA.
In the following, we test a set of different $\delta_0$, $\delta_1$ - combinations and 
evaluate their performance with increasing iteration depth $p$. As a baseline, we consider in particular the case $\delta_0=\delta_1=0.5$ corresponding to a purely classical naive rounding of all variables.

For our simulations, we use as input our generated ensembles of 20 random instances, 20 "hot" and 20 "cold" instances classified by the $\varepsilon$-measure and by the RMSE ($\sigma$-measure) with $N=10$ qubits and $B=5$ selected assets. With this setup, we compute both the approximation ratio $r$ and the probability $P$ for an increasing number of QAOA iterations $p$ (up to $p_{\text{max}}=7$) including the initial state at $p=0$. For the classical preprocessing, we set various bound settings as follows: two symmetric bounds ($\delta_0=\delta_1=0.01$ \text{and} $\delta_0=\delta_1=0.25$), one asymmetric bound ($\delta_0=0.1, \delta_1=0.25$), and the classical baseline as described above ($\delta_0=\delta_1=0.5)$. For comparison, we again plot the results of standard QAOA (without preprocessing) and warm-start QAOA as already displayed above (see Figs.~\ref{fig:ws_std_approx} and \ref{fig:good_bad_eps2}).

\subsubsection{Results for random instances}

\begin{figure*}
	\centering
        \includegraphics[width=0.95\textwidth]{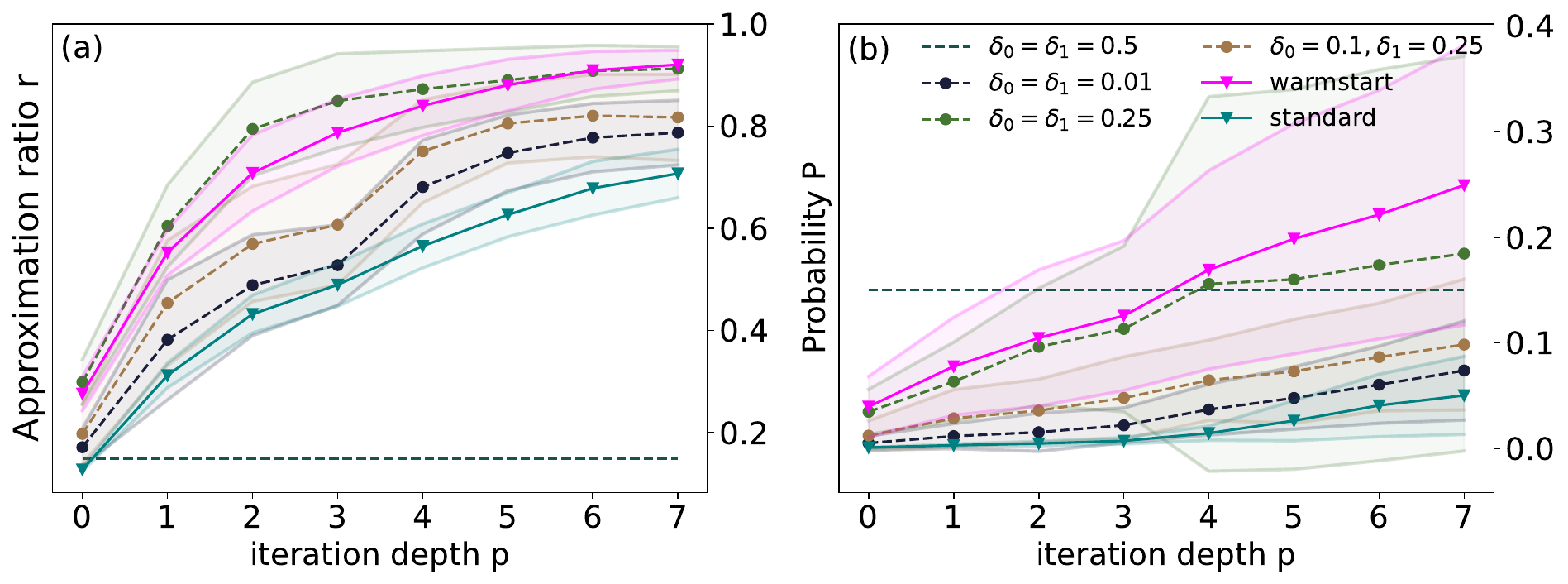}
    \captionsetup{justification=raggedright}
	\caption{Mean approximation ratio (a) and ground state probability (b) obtained with the standard QAOA (teal line), the preprocessed standard QAOA using different bounds (dashed colored lines) and the warm-start QAOA (pink line), for 20 random problem instances with classical baseline (dashed horizontal line). For a bound setting of $\delta_0=\delta_1=0.25$, the improved approximation ratio (a) of warm-start QAOA can be reproduced (or even outperformed at smaller $p$) by the preprocessed standard version. An increase in probability (right) over the standard QAOA is achieved, although not as high as with the warm-start.
 	\label{fig:random_assets}}
\end{figure*}

Fig.~\ref{fig:random_assets} shows the results for the 20 randomly chosen instances already considered in Fig.~\ref{fig:ws_std_approx}.
For the classical baseline ($\delta_0=\delta_1=0.5$), we achieve a mean approximation ratio and mean probability of $r=P=15\%$ ($=\frac{3}{20}$), i.e. 3 out of 20 solutions are in agreement with the optimal solution after the simple rounding (and hence $r=P=1$).
All other cases, where the optimal solution was not found, violate the constraint (thus giving rise to $r=P=0$). Since the classical baseline involves no QAOA optimization, the result does not improve and remains constant over the iteration depth. For the two lower bounds (dashed black and beige lines) we achieve an approximation ratio (left) of about $r \approx 80\%$ at $p=7$ and thus an improvement over the standard QAOA. As can be seen from the initial state, this method provides a better starting point for the algorithm. 

Also concerning the probabilities of obtaining the optimal solution (right), a slight increase over the standard version is visible, but for these three settings (standard and the two lower bounds) the probabilities are below our baseline probability, indicating that the bound settings are not large enough to round the values up or down towards the optimal bits. For a large bound setting (dashed green line) we can reproduce the approximation ratio of the warm-start variant of $r \approx 92\%$ at depth $p=7$ within the error bars, and even surpass it at lower depth. Only the probability of the warm-start is not achievable for the random problem instances. However, compared to the standard QAOA, a significant increase is accomplished. In summary, standard QAOA with preprocessing yields higher approximation ratios than warm-start QAOA, but lower ground state probabilities, indicating that alternative solutions are found that are very close to the optimal solution.

\begin{figure*}
	\centering
        \includegraphics[width=0.95\textwidth]{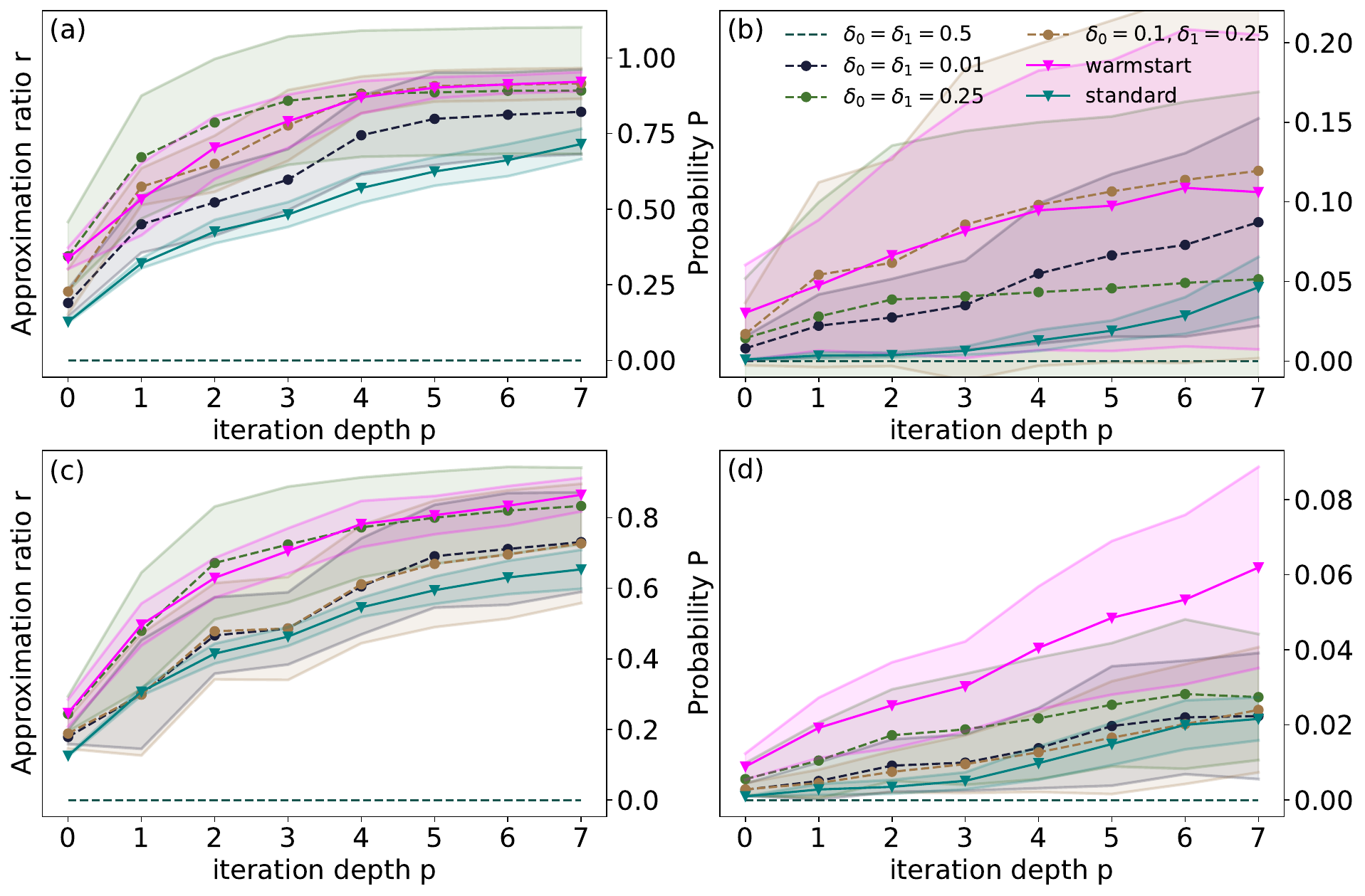}
    \captionsetup{justification=raggedright}
	\caption{Same as in Fig.~\ref{fig:random_assets}, but for $\varepsilon$-cold (a,b) and $\sigma$-cold (c,d) instead of random problem instances. Again, for a large bound setting ($\delta_0=\delta_1=0.25$), the approximation ratio of the warm-start QAOA can be reproduced by the preprocessed standard version for both measures (a,c). However, the ground state probabilities are lower than those obtained with warm-start QAOA (b,d). For a different bound setting ($\delta_0=0.1, \delta_1=0.25$) a higher probability than for the warm-start QAOA can be achieved, but only using instances created with respect to the $\varepsilon$-measure, see (b).}
 	\label{fig:eps_sigma_AR_Prob_bad}
\end{figure*}
\clearpage

\begin{figure*}[htb!]
    \centering
          \includegraphics[width=0.95\linewidth]{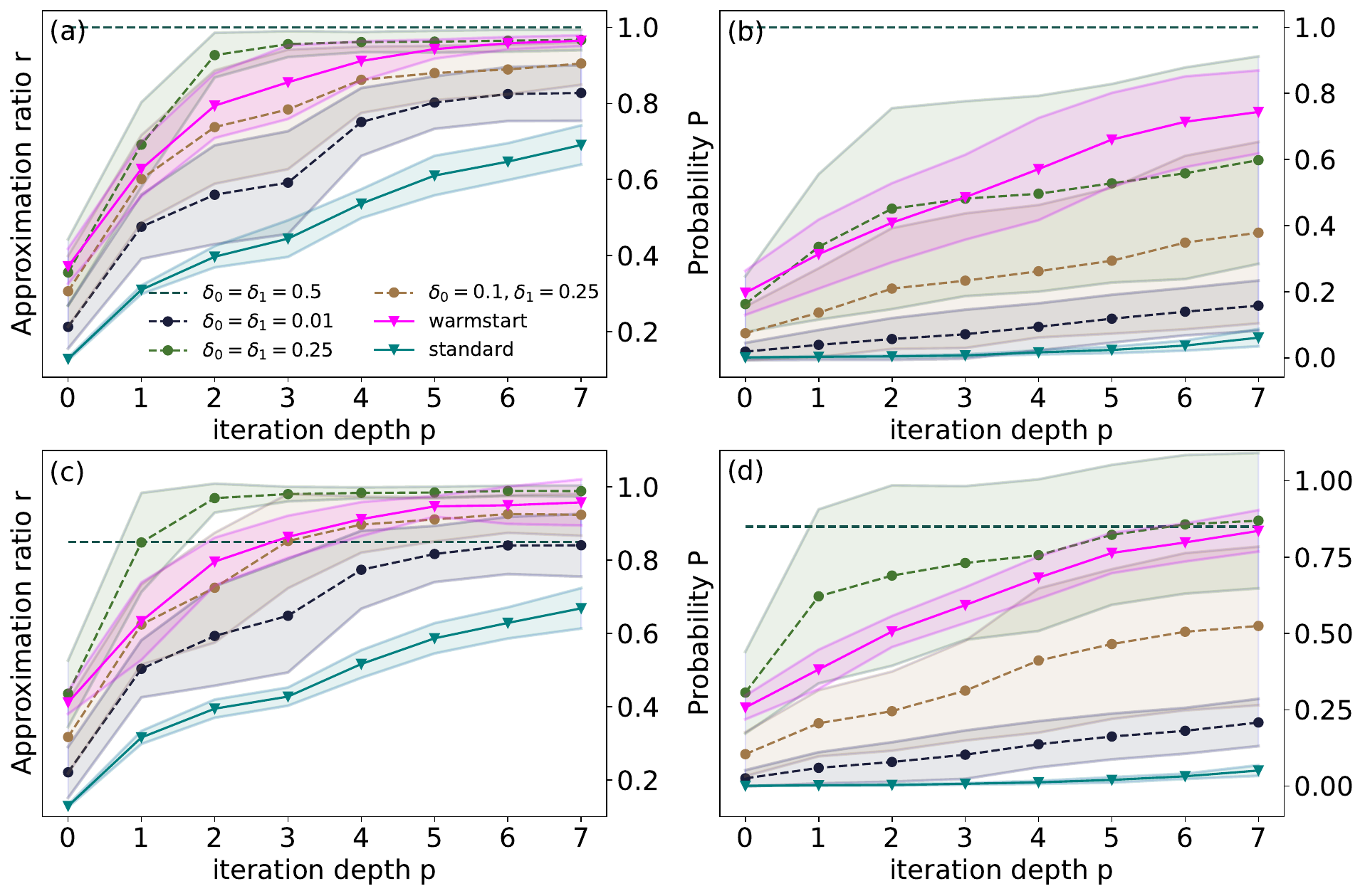}
    \captionsetup{justification=raggedright}
	\caption{Same as in Figs.~\ref{fig:random_assets} and \ref{fig:eps_sigma_AR_Prob_bad} for $\varepsilon$-hot (a,b) and $\sigma$-hot (c,d) problem instances. As for the cold instances, the preprocessed QAOA  with largest bound setting ($\delta_0=\delta_1=0.25$, dashed green line) shows the highest approximation ratio (a,c). In particular, it surpasses warm-start (solid pink line) especially at smaller $p$. In case of the $\sigma$-hot instances, it also displays the largest ground-state probability, see (d). 
 }
 	\label{fig:eps_sigma_AR_Prob_good}
\end{figure*}

\subsubsection{Results for cold instances}

Fig.~\ref{fig:eps_sigma_AR_Prob_bad} shows the performance for the same settings as for the random problem instances, but computed for the "cold" 
instances classified by the $\varepsilon$- and $\sigma$-measures respectively. From the baselines, we see that the optimal solution is never reached by simple rounding of the relaxed solution (since all cold instances exhibit $\varepsilon>0.5$, see Fig.~\ref{fig:scatter}) and the budget constraint is never met. Besides that, the performance of warm-start can be reproduced (approximation ratio of $r_{\varepsilon} \approx 89 \%$ compared to $r_{\varepsilon, \text{WS}} \approx 92\%$ and of $r_{\sigma} \approx 83\%$ compared to $r_{\sigma,\text{WS}} \approx 86\%$ at $p=7$) for both measures in the case of the larger symmetric bound, see the dashed green line in Fig.~\ref{fig:eps_sigma_AR_Prob_bad}(a) and (c). For the $\varepsilon$-measure, see Fig.~\ref{fig:eps_sigma_AR_Prob_bad}(a), it is also achieved with the asymmetric bound (dashed beige line) which, in addition, yields the highest ground state probability, see Fig.~\ref{fig:eps_sigma_AR_Prob_bad}(b). In contrast, for the $\sigma$-measure, the approximation ratios of the lower symmetric and asymmetric bounds, see the dashed beige and black lines in Fig.~\ref{fig:eps_sigma_AR_Prob_bad}(c), coincide within the error bars. Furthermore, the probabilities of finding the optimal solution are below warm-start for all cases, see Fig.~\ref{fig:eps_sigma_AR_Prob_bad}(c).

Note that, in Fig.~\ref{fig:eps_sigma_AR_Prob_bad} (b), we observe extremely large error bars especially in case of the classically preprocessed standard QAOA with bound $\delta_0=\delta_1=0.25$. This is due to the fact that, for most of the 20 $\varepsilon$-cold instances, the ground state probability strictly vanishes, since at least one bit is rounded to the wrong value during the preprocessing. For $\delta_0=\delta_1=0.25$, this occurs if $\varepsilon>0.75$, which, as evident from Fig.~\ref{fig:scatter}, concerns 16 out of the 20 $\varepsilon$-cold instances. In other words, the mean probability originates from only a few instances with $P>0$, which leads to a large standard deviation.

\subsubsection{Results for hot instances}

In Fig.~\ref{fig:eps_sigma_AR_Prob_good}, we now discuss the results for the "hot" problem instances. For these instances, the deviation between the relaxed and the discrete solution is small, so after simply rounding, the optimal solution is expected to be found in most cases. Concerning the $\varepsilon$-measure, see Fig.~\ref{fig:eps_sigma_AR_Prob_good}(a,b), the approximation ratio and probability for the simple rounding (baseline) are both at $r=P=100\%$, reflecting exactly what has just been described. For the other measure ($\sigma$), depicted in Fig.~\ref{fig:eps_sigma_AR_Prob_good}(d,c), the baseline takes an approximation ratio and probability of $r=P=85\%$ ($=\frac{17}{20}$), since 
3 out of the 20 $\sigma$-hot instances exhibit $\varepsilon>0.5$, see Fig.~\ref{fig:scatter}, and violate the constraint after naive rounding.

For the symmetric bound setting with the largest rounding (dashed green line), we obtain a maximum approximation ratio of $r_{\varepsilon} \approx 96,7\%$ and a probability of $P_{\varepsilon} \approx 60\%$ (compared to $r_{\varepsilon,\text{WS}} \approx 96,9\%$ and $P_{\varepsilon,\text{WS}} \approx 77\%$) for the $\varepsilon$-measure, see Fig.~\ref{fig:eps_sigma_AR_Prob_good}(a,b), and an approximation ratio of $r_{\sigma} \approx 98\%$ and a probability of $P_{\sigma} \approx 86\%$ (compared to $r_{\sigma,\text{WS}} \approx 96\%$ and $P_{\sigma,\text{WS}} \approx 83\%$) for the $\sigma$-measure, see Fig.~\ref{fig:eps_sigma_AR_Prob_good}(c,d). With this result, we conclude that, for sufficiently large bound settings, the improved performance of the warm-start QAOA can be reproduced and even be surpassed by standard QAOA with classical preprocessing.

\section{\label{sec:Conclusion}Conclusion and Outlook}

We presented a classical preprocessing approach to warm-start optimization on different problem instances. 
For this purpose, we applied the standard and the warm-start version of the QAOA algorithm to the portfolio optimization problem using an ensemble of 20 random, "hot", and "cold" instances with $N=10$ assets, where the two latter ones are created by comparing the relaxed and discrete solution of 1000 random instances in terms of the maximum deviation per instance $\varepsilon$ and the root mean square error $\sigma$. 

Concerning the performance of the standard QAOA, we have found that dividing into the two subsets does not show an improved or worse performance compared to using the random instances. In contrast, for the warm-start QAOA, we clearly see that the "hot" instances are better to optimize than the "cold" instances, although an advantage as compared to standard QAOA still holds for the latter ones.

We introduced a classical preprocessing approach to warm-start optimization by first using a rounding scheme on the relaxed solution of the continuous problem for values close to 0 or 1 and then solving the reduced problem with standard QAOA. As a result, the performance of the standard QAOA, in terms of the approximation ratio, can be boosted by applying this classical preprocessing and, especially for smaller $p$, it also outperforms the results of warm-start QAOA if the bounds for the rounding process are large enough. For the ground state probability, we observe an increase compared to the values generated by the standard QAOA, but, in general, we find them to be lower than those of warm-start QAOA.  

Finally, we conclude that the improved performance of warm-start can be reproduced by classical methods and is thus not a result of quantum effects alone. In future work, this insight could be useful in order to explore further ways of splitting an optimization routine into classical and quantum processing parts and thereby to realize a quantum advantage over purely classical methods. A better understanding of which steps in the optimization routine can be replaced classically will be helpful to achieve a better performance through more targeted use of quantum computing.

\section{Aknowledgement}
This work is funded by the Ministry of Economic Affairs, Labour and Tourism Baden
Württemberg in the frame of the Competence Center Quantum Computing Baden-Württemberg (project ‘QORA II’).

\appendix

\section{Portfolio data}

In the tables \ref{tab:return} and \ref{tab:covariance}, the return vector and covariance matrix are given as an example for one of the random portfolio instances used in section \ref{sec:warm_start_instances}. Other instances can be generated as described in \cite{Brandhofer2022} (supplementary material). \\

\onecolumngrid

\begin{table*}[h]
\resizebox{\linewidth}{!}{
\begin{tabular}{lllccccrrr}
\specialrule{.1em}{.05em}{.05em}
FRE.DE & DTE.DE & IFX.DE & SIE.DE & ALV.DE & BAS.DE & HEN3.DE & LIN.DE & RWE.DE & MUV2.DE\\
\specialrule{.1em}{.05em}{.05em} \\
-0.07998594 & 0.0444345 & 0.20639829 & 0.10283742 &  0.1030686 &  0.05094806 & 0.00832845 & 0.26801758 & 0.30300314 & 0.1128935\\
\hline 
\end{tabular}
}
\caption{Return vector $\mu_{i}$ for 10 assets chosen from the German DAX 30}
\label{tab:return}
\end{table*}

\newcolumntype{?}{!{\vrule width 0pt}}

\begin{table*}[h]
\resizebox{\linewidth}{!}{
\begin{tabular}{l?lllccccrrr}
 & FRE.DE & DTE.DE & IFX.DE & SIE.DE & ALV.DE & BAS.DE & HEN3.DE & LIN.DE & RWE.DE & MUV2.DE\\
\specialrule{.1em}{.05em}{.05em} \\
FRE.DE & 0.0887543 & 0.02804199 & 0.05163708 & 0.0414294 & 0.04357646 & 0.04001491 & 0.02689946 & 0.02367172 & 0.03149689 & 0.03734508 \\
DTE.DE & 0.02804199 & 0.03999579 & 0.03292228 & 0.02906106 & 0.03254393 & 0.02991219 & 0.02124013 & 0.01764212 & 0.02967336 & 0.03081729 \\
IFX.DE & 0.05163708 & 0.03292228 & 0.13060002 & 0.05737477 & 0.0550938 & 0.05734743 & 0.0311434 & 0.04352947 & 0.04571444 & 0.04903502 \\
SIE.DE & 0.0414294 & 0.02906106 & 0.05737477 & 0.06472856 & 0.0489978 & 0.05052971 & 0.02730231 & 0.02944869 & 0.03578296 & 0.04289097 \\
ALV.DE & 0.04357646 & 0.03254393 & 0.0550938 & 0.0489978 & 0.06531374 & 0.0503312 & 0.02770636 & 0.02860439 & 0.03487315 & 0.05488462 \\ 
BAS.DE & 0.04001491 & 0.02991219 & 0.05734743 & 0.05052971 & 0.0503312 & 0.06984039 & 0.02911981 & 0.02908292 & 0.03458258 & 0.04606973 \\
HEN3.DE & 0.02689946 & 0.02124013 & 0.0311434 & 0.02730231 & 0.02770636 & 0.02911981 & 0.04272304 & 0.01806647 & 0.02791279 & 0.02735153 \\
LIN.DE & 0.02367172 & 0.01764212 & 0.04352947 & 0.02944869 & 0.02860439 & 0.02908292 & 0.01806647 & 0.21117209 & 0.02104146 & 0.02922818 \\
RWE.DE & 0.03149689 & 0.02967336 & 0.04571444 & 0.03578296 & 0.03487315 & 0.03458258 & 0.02791279 & 0.02104146 & 0.11322095 & 0.03446745 \\
MUV2.DE & 0.03734508 & 0.03081729 & 0.04903502 & 0.04289097 & 0.05488462 & 0.04606973 & 0.02735153 & 0.02922818 & 0.03446745 & 0.06765634 \\
\specialrule{.1em}{.05em}{.05em}
\end{tabular}
}
\caption{\label{tab:covariance} Covariance matrix $\sigma_{ij}$ for 10 assets chosen from the German DAX 30}
\end{table*}
\twocolumngrid

\newpage
\bibliography{mybibliography.bib}
\end{document}